\input harvmac
\newcount\figno
\figno=0
\def\fig#1#2#3{
\par\begingroup\parindent=0pt\leftskip=1cm\rightskip=1cm
\parindent=0pt
\baselineskip=11pt
\global\advance\figno by 1
\midinsert
\epsfxsize=#3
\centerline{\epsfbox{#2}}
\vskip 12pt
{\bf Fig. \the\figno:} #1\par
\endinsert\endgroup\par
}
\def\figlabel#1{\xdef#1{\the\figno}}
\def\encadremath#1{\vbox{\hrule\hbox{\vrule\kern8pt\vbox{\kern8pt
\hbox{$\displaystyle #1$}\kern8pt}
\kern8pt\vrule}\hrule}}

\overfullrule=0pt

\Title{\vbox{\baselineskip12pt
\hbox{hep-th/9911137}
\hbox{TIFR-TH/99-57}
\hbox{MRI-PHY/P99113}
\hbox{IMSc/11/37}
}}
{\vbox{\centerline{Supergravity with Self-dual $B$ fields and}
\centerline{Instantons in Noncommutative Gauge Theory}}}
\smallskip
\centerline{Sumit R. Das~$^{a,b}$\foot{das@theory.tifr.res.in},
S. Kalyana Rama~$^c$ \foot{krama@imsc.ernet.in} and Sandip P. Trivedi~
$^a$ \foot{sandip@theory.tifr.res.in}}
\smallskip
\centerline{ $~^a$ {\it Tata Institute of Fundamental Research}}
\centerline{\it Homi Bhabha Road, Bombay 400 005, INDIA}
\smallskip
\centerline{$^b${\it Mehta Research Institute of Mathematics and 
Mathematical Physics}}
\centerline{{\it Jhusi, Allahabad 211 019, INDIA}}
\smallskip
\smallskip
\centerline{$^c${\it Institute of Mathematical Sciences}}
\centerline{\it C.I.T. Campus, Chennai 600113, INDIA}
\bigskip

\medskip

\noindent
We study Type IIB supergravity in the presence of (euclidean) D3
branes and nonzero self-dual $B$-fields.  We point out that the
Einstein frame metric is identical to the full geometry for D3 branes
without B fields turned on.  Furthermore, in a decoupling limit in
which the theory is conjectured to be dual to noncommutative
Yang-Mills theory, the entire Einstein metric remains intact,
and in particular, is asymptotically flat.  We
construct D-instanton solutions in this geometry.  We show that in the
decoupling limit the D-instanton action agrees with the action of the
corresponding instanton in the noncommutative Yang-Mills theory and
is expressed in terms of the open string coupling.  Some other aspects
of this correspondence, which have unusual features because the
underlying metric is asymptotically flat, are explored.

\Date{November, 1999}

\def\tB{{\tilde B}}
\def\sqg{{\sqrt{g}}}
\def\tH{{\tilde{H}}}
\def\bphi{{\bar{\phi}}}
\def\bchi{{\bar{\chi}}}
\def\bx{{\bar{x}}}
\def\bu{{\bar{u}}}
\def\btheta{{\bar{\theta}}}
\def\sqgs{{\sqrt{g_s}}}
\def\tg{{\tilde{g}}}

\def\tx{{\tilde{x}}}
\def\hg{{\hat{g}}}

\newsec{Introduction}

The observation that the low energy dynamics of D-branes in the presence
of constant NSNS B fields is noncommutative Yang-Mills theory
\ref\douglashull{A. Connes, M. Douglas and A. Schwarz,
JHEP 9802 (1998) 003, hep-th/9711162; M. Douglas and C. Hull, JHEP 02
(1998) 008, hep-th/9711165.} has opened up the possibility of
exploring string theory in a new limit.  In this limit the string
tension is scaled to infinity, and at the same time the
closed string metric is scaled to zero, 
 keeping the dimensionful NSNS B field
fixed \ref\seibergwitten{N. Seiberg and E. Witten, JHEP 9909:032
(1999), hep-th/9908142.}. This is different from the usual low
energy limit where the closed string metric and the dimensionless NSNS
B-field are kept fixed as the tension scales to infinity.  As a
result, the lowest order effective action on the brane contains an
infinite number of terms with all powers of the field strength when
written in terms of ordinary gauge fields and conventional gauge
symmetry. However, as shown in \seibergwitten, in terms of
noncommutative gauge fields, one has a simple lagrangian. 
The fact that an infinite number of terms may be
neatly summarized in terms of a different geometric structure raises
hopes that one may be able to perform concrete calculations in this
new corner of string theory. Various aspects of noncommutative gauge
theory and its relationship with string theory have been studied in
\ref\noncom{Y.K.E. Cheung and M. Krogh, Nucl. Phys. B528 (1998) 185,
hep-th/9803031; C.S. Chu and P.M. Ho, Nucl. Phys. B550 (1999) 151,
hep-th/9812219;
F. Ardalan, H. Arfaei and M.M. Sheik-Jabbari, JHEP 02:016 (1999),
hep-th/9810072;  V. Schomerus, JHEP 9906:030 (1999), hep-th/9903205;
D. Bigatti and L. Susskind, hep-th/9908056.}.

Over the past two years it has become clear that large-N
supersymmetric Yang-Mills theories are holographic descriptions of
string theories in suitable backgrounds,
\ref\adscft{J. Maldacena, Adv.Theor. Math. Phys. 2 (1998) 231, 
hep-th/9711200; S. Gubser, I. Klebanov and A. Polyakov,
Phys. Lett. B428 (1998) 105,hep-th/9802109; E. Witten,
Adv. Theor. Math. Phys. 2 (1998) 253, hep-th/9802150}
one of the best understood example being the duality between large-N
${\cal N}=4$ Yang-Mills theory in four space-time dimensions and $IIB$
string theory on $AdS_5 \times S^5$. When the 't Hooft coupling of the
Yang-Mills theory is strong the dual string theory reduces to
supergravity. The Yang-Mills theory in question is the low energy
theory of $N$ three branes and the $AdS_5 \times S^5$ background of
the supergravity is the near-horizon limit of the three-brane
background. It is natural to expect that in the presence of a $B$ field,
one would have a similar correspondence between noncomuutative
Yang-Mills theory and supergravity backgrounds in the presence of
suitable $B$ fields. Indeed in \ref\hashimoto{A.
Hashimoto and N. Itzhaki, hep-th/9907166.} and \ref\maldarusso{J.
Maldacena and J. Russo, hep-th/9908134.} dual 
supergravity backgrounds for noncommutative boundary theories were 
proposed and certain aspects of holography were investigated in
\maldarusso\ and \ref\miaoli{M. Li and Y.S. Wu, hep-th/9909085.}.
Other examples of such backgrounds have recently appeared in
\ref\mikhailov{M. Alishahiha, Y. Oz and M. Sheikh-Jabbari,
hep-th/9909215; A. Mikhailov, hep-th/9910126; A. Dhar, G. Mandal, S.R. Wadia
and K.P. Yogendran, hep-th/9910154.}.

In this paper we will continue the study of the supergravity
background dual to $3+1$ dimensional noncommutative $SU(N)$ Yang Mills
theory.  There are two motivations for this study. First, as in the
ordinary case, one might hope to learn something about noncommutative
Yang Mills theory at large 'tHooft coupling in this way. Second, and
perhaps more interestingly, one might hope to learn about holography
in contexts other than those of AdS space.

In particular we will focus on the case where the two form $B$ fields are 
self dual. 
  In this case as was discussed in \seibergwitten\ 
the instanton moduli space in the noncommutative Yang Mills theory 
 is the same as that
of ordinary Yang Mills theory and in particular, the size of the instanton
 is one of the moduli. 
Correspondingly, the D-instanton is a BPS state state in the theory 
and one expects to be able to localize it anywhere in the bulk supergravity
geometry. 
 The situation is therefore similar to that of D-instantons
in AdS space \ref\dinstantons{C.S. Chu, P.M. Ho
and Y.Y. Wu, hep-th/9806103; M. Bianchi, M. Green, S. Kovacs and
G. Rossi, hep-th/9807033} where 
the location of the D-instanton explicitly 
maps to the scale size of the Yang Mills
instanton and one point functions in the D-instanton background
accurately  reproduce the Yang Mills instanton profile \ref\trivedi{V.
Balasubramanium, P. Kraus, A. Lawrence and S.P. Trivedi,
hep-th/9808017.}. 
The hope is that in the non-commutative case as well the study of instantons
for the self dual case would be revealing in the study of the IR/UV relation
and more generally holography\foot{Of course instantons in the 
general case when $B^{-} \ne 0$ are also interesting, perhaps even more so,
since in that case the moduli space is different from the ordinary Yang Mills
case and in particular the small instanton singularity is absent 
\ref\nekarasov{
N. Nekrasov and A. Schwarz, Comm. Math. Phys. 198 (1998) 689,
hep-th/9802068} . However,
 since these are not threshold
bound states the supergravity solutions are harder to find.}.

Another, and perhaps deeper, reason for being interested in the
supergravity background for the case $B^-=0$ comes from attempts to
understand holography in flat space. As we will see below, the 
Einstein frame metric for self dual $B$ field configurations,
 in the scaling limit which corresponds to noncommutative Yang Mills
theory, is {\it the full} three brane geometry in the absence
of a $B$ field.  This remarkable fact has several implications. The
most noteworthy being that the geometry with a self dual
$B$ field is in fact asymptotically flat.  We should add that other
fields in particular the dilaton and axion are not constant. Even so,
this fact suggests that ultraviolet noncommutative Yang Mills
theory should help in understanding holography in flat $10$-dim.
space. It has been suggested in \ref\ghaskl{S. Gubser, A. Hashimoto and
I. Klebanov, Nucl. Phys. B 526 (1998) 393, hep-th/9803023.}
that in the case without B fields
the full D3 brane geometry corresponds to keeping the
Born-Infeld action in the Yang Mills theory and not truncating to
the lowest dimension operator. Here we see that the effect of exciting
the other fields in supergravity- dilaton-axion and two forms-
corresponds in the Yang Mills theory to a different choice of higher dimension
operators -those which can be organised into the 
 noncommutative Yang Mills Lagrangian.

This paper is structured as follows. In section 2, we review the supergravity
solution for a self dual $B$ field and in particular show that its near
horizon limit gives rise to the full three brane geometry without
a B field turned on. In section 3, we then construct the D-instanton solution.
Because of the remarks above, this is also a solution for a 
D-instanton in the full D3 brane geometry without a B field. 
Section 4, discusses the supersymmetries of the solution,  and  finally
section 5, some aspects of holography.

\newsec{Supergravity Solutions with self dual B fields}
\leftline{\bf 2.1 The backgound geometry}
We will find euclidean solutions to the IIB supergravity equations of
motion which are listed in Appendix I. In our notation, 
the NS fields will be real
while the RR fields will be purely imaginary.  Start with the
following ansatz, for the dilaton $\phi$, RR scalar $\chi$, and NS and RR
two form potentials $B_{\mu\nu}$ and $\tB_{\mu\nu}$.
\eqn\fourt{\chi +i~\gamma~ e^{-\phi} = ic}  
\eqn\four{\eqalign{
& \tH_{\mu\nu\alpha} = ic H_{\mu\nu\alpha}\cr
& B_{01} = -\gamma B_{23}~~~~~~~\tB_{01} = - \gamma \tB_{23}}}
where $c$ is a real constant. Here $H$ and $\tH$ denote the field strengths 
for $B$ and $\tB$. This ansatz  
is motivated by supersymmetry in the following
section.

With this ansatz, the energy momentum tensors of $\phi$ and $\chi$
cancel each other and so does those of $B$ and $\tB$ so that 
Einstein equations (7.1) become
\eqn\five{R_{\mu\nu} - {1\over 2}g_{\mu\nu}R = T_{\mu\nu}^F,}
where $T_{\mu \nu}^F$ denotes the stress energy of the five form field strength.
The five form in turn is determined by Gauss' law
\eqn\glax{{1 \over \sqrt{g}}\partial_\mu(\sqrt{g}
F^{\mu \lambda \alpha \beta \gamma})=0.}
Finally,
the dilaton and the axion equations ((7.2) and (7.3) respectively)
both reduce to the following equation
\eqn\six{\nabla^2 (e^\phi) + {1\over 6} H_{\mu\nu\alpha}H^{\mu\nu\alpha}
= 0}
while the $B$ and the $\tB$ equations ((7.4) and (7.5) respectively)
both become
\eqn\seven{{1\over \sqg}\partial_\mu(\sqg H^{\mu\nu\lambda})
- {2\over 3}i ~F^{\nu\lambda\alpha\beta\gamma}H_{\alpha\beta\gamma}
= 0.}

The important point is that with  the ansatz, \four\ \fourt\ the equations for
the metric \five\ and the $5$-form, \glax\ form a closed set
{\it independent} of the two-forms and the dilaton and  axion fields.
Thus if a solution consistent with the ansatz \four\ exists
it give rise to a metric and $5$-form identical to the
case of the $D3$ brane solution {\it without} any two-form fields
turned on.

One of the supergravity backgrounds presented in \maldarusso\ is in
fact a solution of the equations of motion with the ansatz \four.
The metric (in the Einstein frame)
  is given  \foot{The solution below differs from that given
in \maldarusso\ by a relative negative sign between the RR and the
NS-NS 2-form fields. We have checked explicitly that the following solves
the equations of motion.}
by 
\eqn\defmetric{ds_E^2={1\over \sqgs}
\bigl [~ [1+{4 \pi g_s N (\alpha ')^2 \over r^4}]^{(-1/2)}
(dx_0^2 + \cdots + dx_3^2) ~+~
[1+{4 \pi g_s N (\alpha ')^2 \over r^4}]^{(1/2)}(dr^2 + r^2 d\Omega_5^2) ~\bigr ]}
The five form is
\eqn\deffive{F_{0123r}=({4 i \over g_s})~ \bigl [
 {(4 \pi g_s N)^2 (\alpha^{'})^2
\over (1 + {4 \pi g_s N (\alpha^{'})^2 \over r^4})^2} \bigr ]~ { 1 \over r^5},}
and the dilaton, axion and two -forms are 
\eqn\elevena{\eqalign{
&e^{-\phi_0} =  {\cos^2\theta\over g_s} - i~\chi_0 = {1\over g_s~h(r)}\cr
& \tB_{01} = \tB_{23} = - {i\cos^2\theta\over g_s} B_{01} =
-{i\cos^2\theta\over g_s} B_{23}
= -{i\over g_s}\sin\theta \cos\theta {h(r)\over f(r)}}}
where
\eqn\twelve{\eqalign{&f(r) = 1+ {\alpha'^2 R^4 \over r^4}\cr
&h^{-1}(r) = \sin^2\theta f^{-1}(r) + \cos^2\theta}}
The parameter $\theta$ is related to
the $B$-fields at $r = \infty$, $g_s$ is the asymptotic value of the string
coupling and
\eqn\defR{R^4 = {4\pi g_s N\over \cos^2\theta}.}
This is a solution of the equations of motion with the ansatz \four\
with $c = -{\cos^2\theta\over g_s}$ and $\gamma = -1$.  A similar
solution can be of course written down with $\gamma = 1$.
In particular we note that the metric and five-form are independent
of $\theta$ and equal to their values for a $D3$ brane geometry without
any $B$ fields turned on. 

Now one can take a scaling limit which corresponds to the low-energy limit
of \seibergwitten\ that gives rise to noncommutative Yang Mills theory. 
This is given by \maldarusso:
\eqn\twoonej{\alpha' \rightarrow 0 ~~~~~\theta \rightarrow 
{\pi \over 2}}
with
\eqn\twoonel{ {\tilde b} = \alpha'~\tan\theta,~~\tg = {g_s \over (\alpha')^2}
~~{\rm and} ~~r/\alpha'}
held fixed. 

The reader will notice that the harmonic function appearing in the
metric cannot be not truncated to its near horizon value \foot{This is
because the string coupling is also being scaled to zero.}.  Instead,
as was mentioned in the introduction, the full $D3$ brane metric
survives in this limit.

The full solution in this limit  can be written in terms of the variable
\eqn\defu{ u= r/(\alpha' R^2)}
used in \maldarusso. After the rescaling coordinates $x_i$
along the brane as:
\eqn\twotwoj{x_i \rightarrow \tx_i = {{\tilde b} \over \alpha'}x_i
~~~~~~~~i=0,1,2,3}
the solution becomes
\eqn\eleven{\eqalign{& ds^2 = {\alpha ' R^2 \over \sqrt{\hat g}}
[(f(u))^{-1/2}(d\tx_0^2 + \cdots + d\tx_3^2) +
(f(u))^{1/2} (du^2 + u^2 d\Omega_5^2)]\cr
&e^{-\phi_0} =  {1\over \hg} - i~\chi_0 = {1\over \hg} u^4 f(u)\cr
& \tB_{01} = \tB_{23} = -{i\over \hg} B_{01} = -{i\over \hg} B_{23}
= - {i \alpha ' a^2 R^2 \over \hg}~(f(u))^{-1}\cr
&F_{0123u} = {4i(\alpha ')^2 R^4\over \hg u^5}(f(u))^{-2}}}
where
\eqn\twelve{f(u) = {1\over u^4} + a^4,}
and
\eqn\twelvedefa{a^2={\tilde b} R^2.}
The coupling has been also redefined
\eqn\defhg{\hg= {\tilde b}^2 {\tilde g}}
while $R^4$, defined in
\defR\   can be expressed as
\eqn\twelveg{R^4 = 4\pi \hg N,}

Note that the coupling $\hg$ is precisely the open string coupling
which appears in noncommutative Yang-Mills theory in the low
energy limit defined in \seibergwitten.

The metric \eleven\ becomes $AdS_5 \times S^5$ near the horizon,
$u=0$ and $\hg$ is the value of the string coupling in the IR. 
The $B$ fields vanish at $u=0$ and asymptote to $\alpha'/{\tilde b}$
as $u \rightarrow \infty$. 

It is important to note that the scaling limit considered above is
{\it not} the near-horizon limit of the full geometry. This is the
supergravity reflection of a corresponding feature of the low energy
limit of the D3 brane theory considered in \seibergwitten, viz. the
fact that arbitrarily high powers of the (ordinary) field strengths
survive.
 
In the rest of the paper we will work with the variables of \eleven
and rename ${\tilde x} \rightarrow x$. 

\leftline{\bf  2.2 D-instanton Solution}
We now turn to determining the D-instanton solution in the background
\defmetric. We have already emphasized that \five and \glax\ are independent
of the two-forms and the dilaton-axion fields. 
In addition, the equation \seven\
for the three-form field strength does not involve the dilaton or axion
fields either. Denoting the dilaton and axion fields appearing in 
\defmetric\ as $\phi_0, \chi_0$ it then follows that a solution different 
from \elevena\ can be obtained with the same Einstein metric, two form
fields and five form field strength but with different 
dilaton and axion values $\phi$ and $\chi$
\eqn\eight{e^\phi = e^{\phi_0} + e^{\bphi}~~~~~\chi = \chi_0 + \bchi}
so that 
\eqn\nine{i\chi +~ e^{-\phi} = {1\over {\hat g}}}
and
\eqn\ten{\nabla^2 (e^\bphi) = 0}

We are now ready to write down the equation satisfied by the fields
due to a D-instanton. The metric, $B$-fields 
and the five form field strength are exactly the same as in \elevena.
Since the D-instanton couples to the dilaton and axion, the equation 
satsified by
$\bphi$ is modified from \ten\ at the location of the instanton. 

The D-instanton action is given by,
\eqn\actd{S_{I}=2\pi ~ \int [e^{-\phi}+i\chi]\delta^{10}(x-\bx) d^{10}x.}
Here  $\bx = (\bx^0 \cdots \bx^3, \bu,
\btheta_i)$ is 
the location of the D-instanton 
(with $\theta_i$ denote the angles parametrizing the $S^5$). Thus the equation 
for the 
dilaton field is given by
\eqn\thrirteen{e^{\phi (x,\bx)} = e^{\phi_0 (u)} + e^{\bphi (x,\bx)}}
where
\eqn\fourteen{\nabla_x^2 (e^{\bphi (x,\bx)}) = - 4 \pi \kappa_{10}^2
 {\delta^{10} (x-\bx)
\over \sqg}.}
In other words we have to solve for the scalar Green's function in the
metric given in \elevena. 

This establishes the existence of the D-instanton solution in the
presence of D3 branes and a nonzero B field. In the following section we
will find explicit solutions to \fourteen\ in the decoupling limit,
i.e in the background given by \eleven.

Before doing so let us make one comment regarding the instanton action
\actd. We started by an ansatz for solutions to supergravity equations.
This ansatz can be in fact be motivated by the action \actd. 
As was mentioned in the introduction, we expect in the 
self dual case that the three brane geometry should allow 
for a D instanton localised anywhere in the bulk. 
This means that the
D-instanton action must be a constant. From \actd\ we see that this implies
the condition \fourt.  A similar argument then also shows that 
the full instanton and background solution satisfies \nine.
In the more general case where $B^- \ne 0$, and the instanton moduli
space is deformed from the ordinary case, \actd\ will yield the 
potential generated in moduli space in the large 'tHooft coupling limit \foot{
We thank D. Tong for a conversation on this point.}.

\newsec{The smeared D-instanton solution}

We will now find the solution of \fourteen\ in the decoupling limit
which is smeared over
the $S^5$. In other words we look for solutions of the equation
which do not depend on the angles $\theta_i$. This would give us the
field due to a D-instanton which is smeared over the five sphere -
the field would be then independent of $\theta_i$. Furthermore,
using translation invariance along the brane directions
$x^0 \cdots x^3$ this solution may be written as
\eqn\fifteen{e^\bphi = {4 \kappa_{10}^2 \hg^2 
\over \pi^2 (\alpha^{'} R^2  )^4} ~
 \int d^4 k { 1\over (2 \pi)^4} ~e^{-ik \cdot (x-\bx)}~G_k(u,\bu),}
where $G_k (u,\bu)$ now solves
\eqn\sixteen{\partial_u (u^5 \partial_u G_k)
- u(1 + a^4 u^4)~k^2~ G_k = - \delta (u - \bu)}
Note that we are looking for euclidean solutions so that $k^2 > 0$.
Also  $G_k$ should be regular as $u \rightarrow 0$,
and  $G_k \rightarrow 0$ when $u \rightarrow \infty$. 
The normalisation in \fifteen\ is obtained from \fourteen\ by integrating over 
the $S^5$ and noting that the volume of the unit five sphere is $\pi^3$.

The Green's function can be expressed in terms of solutions to the 
homogeneous equation,
\eqn\sixteena{\partial_u (u^5 \partial_u G_k)
- u(1 + a^4 u^4)~k^2~ \psi_k = 0,}
by standard Strum-Liouville theory techniques.  
\sixteena\ was studied in 
 \ref\gubserhashimoto{S. Gubser and A. Hashimoto,
Comm. Math. Phys. 203 (1999) 325, hep-th/9805140} and subsequently 
also encountered in \maldarusso, for computing the two point function of
 a particular supergravity mode. 
Solutions to \sixteena\ can be 
written as 
\eqn\defpsi{\psi_k(\nu,u)={1 \over u^2} \phi_k(\nu,u),}
where $\phi_k(\nu,w)$ solves the Mathieu equation, 
\eqn\nineteen{[\partial_w^2 + 2 q \cosh 2w - 4]\phi_k (\nu,w) = 0}
with
\eqn\eighteen{ u = {1\over a} e^{-w},}
and,
\eqn\nineteena{q = - (ka)^2.}
The parameter $\nu$ which enters the solution
is determined by recursion relations
and may be expressed as a power series in $(ka)$ as \gubserhashimoto\
\eqn\twenty{ \nu = 2 - {i{\sqrt {5}}\over 3}({{\sqrt{q}}\over 2})^4
+ {7i \over 108 {\sqrt{5}}}({{\sqrt{q}}\over 2})^8 + \cdots}

In order to implement the boundary conditions on the Greens function
we need two solutions to \sixteena\ one being  regular at $u \rightarrow 0$
and the second going to zero as $u \rightarrow \infty$. 
In the asymptotic region, $|w| \gg 1$,
solutions to the Mathieu equation  
can be expressed in terms of modified Bessel functions. 
The two required solutions can then be written down as :
\eqn\twoone{\eqalign{
&\psi^{(1)}_k (\nu,u) = {i\pi \over 2 u^2}e^{{i\pi \nu \over 2}}
~H^{(1)} (\nu,w) \cr
&\psi^{(2)}_k (\nu,u) = {i\pi \over 2 u^2}e^{{i\pi \nu \over 2}}
~H^{(1)} (\nu,-w) }.}
Here, we are using the notation of \gubserhashimoto\
and   $H^{(1)}(\nu,w)$ denotes a solution which asymptotically
is related to the modified Bessel function, $K_{\nu}$, i.e.,
\eqn\twotwo{ 
{i\pi \over 2} e^{i\pi \nu \over 2} H^{(1)}(\nu,w)  
= K_\nu (ka~e^w) \rightarrow 
\sqrt{\pi \over 2 ka e^w }~ e^{-(ka e^w)} ~~~~~~w \gg 1}. 
Thus we see that 
$\psi_k^{(1)}(\nu,u)$
is well behaved at $u = 0$ while  $\psi_k^{(2)}(\nu,u) \rightarrow 0$
 at $u = \infty$, as required. 

Now using Strum Liouville theory the Greens function is given by:
\eqn\twofour{\eqalign{& G_k (u,\bu) = 
A^2(k,a)~\psi_k^{(1)}(\nu,\bu)~
\psi_k^{(2)}(\nu,u)~~~~~~~~\bu < u\cr
& G_k (u,\bu) = A^2(k,a)~\psi_k^{(1)}(\nu,u)~
\psi_k^{(2)}(\nu,\bu)~~~~~~~~\bu > u}}
The normalization $A^2(k,a)$ has to be determined by requiring that
\eqn\norm{u^5 W(\psi_k^{(1)}, \psi_k^{(2)}) A^2(k,a) =-1,}
where $W(\psi_k^{(1)}, \psi_k^{(2)})$ denotes the Wronskian of the 
two solutions. More detailed forms of the asymptotic behavior of these
functions and the calculation of the Wronskian is given in 
Appendix II.

The result for the Wronskian 
fixes the normalization appearing in \twofour\ to be
\eqn\threetwo{A^2(ka) = {2\sin \pi\nu \over \pi}
(\eta \zeta -{1\over \eta\zeta})^{-1}}
where $\zeta$ is defined in Appendix II.

We now have an expression for the solution to \fourteen
\eqn\threethree{e^\bphi =  {4 \kappa_{10}^2 \hg^2 \over \pi^2 (\alpha^{'} R^2
  )^4}
\int [d^4k] {1\over (2 \pi)^4} ~e^{-ik(x-\bx)}
~A^2(ka)~\psi_k^{(1)}(\nu,\bu)~\psi_k^{(2)}(\nu,u)~~~~~~~u > \bu}
The parameter $\nu$ and the function $\phi(\nu)$ (introduced in
Appendix II) are explicitly
defined in terms of the parameters in the Mathieu equation.
However we will be able to extract some of the physics by
knowing asymptotic forms. In Appendix II we derive the small $ka$
behavior of the various functions which appear in the solution
to Mathieu's equation. In particular it is shown that
\eqn\threethreea{A^2(ka) \sim (ka)^4~~~~~~~~(ka) << 1}

Of particular interest is the small $(ka)$, fixed $(au)$ behavior of the
solutions. We reproduce the results from Appendix II
\eqn\xtwo{\eqalign{&\psi_k^{(1)}(\nu,u) \rightarrow {1\over u^2}K_2(k/u)\cr
&\psi_k^{(2)}(\nu,u) \rightarrow {1\over u^2}K_2(ka^2u)}}
This, together with \threethreea\ shows that the zero momentum
Green's function is
\eqn\sixfour{\eqalign{& G_k (u,\bu)|_{k=0} \sim {1\over u^4}~~~~~~~~~u>\bu\cr
& G_k (u,\bu)|_{k=0} \sim {1\over \bu^4}~~~~~~~~~u < \bu}}
This is what is expected from the differential equation definining
$G_k$ ,equation \sixteena\ and is in fact the zero momentum Green's function
in $AdS_5$ as well. Corrections to this extreme limit may
be worked out systematically.

This shows that it is useful to recast \threethree\ in a somewhat different
form. 
The zero momentum component of $\psi_{k=0}^{(2)}(\nu,u)$ decays 
for large $u$ not exponentially but as a power law. It is useful for 
various purposes to show this dependence explicitly.
Thus we can write:
\eqn\exform{\eqalign{e^{\bphi}= {4 \kappa_{10}^2 \hg^2 \over \pi^2 
(\alpha^{'} R^2  )^4}
\bigl[{1\over 4 L^4 u^4} + \int [d^4k] {1\over (2 \pi)^4} ~e^{-ik(x-\bx)}
~A^2(k,a)~\psi_k^{(1)}(\nu,\bu)~&\psi_k^{(2)}(\nu,u)\bigr ]\cr
&u > \bu}}  
where the first term is the zero momentum  piece, with $L^4$
being the volume along the four directions parallel to the $3$ branes. 
The integral in \exform\ is understood to be over the non-zero momentum
modes.  

It may be checked that the solution \exform\ reproduces known limits.
When both $u$ and $\bu$ are small and at small momenta we should
recover the D-instanton in $AdS_5 \times S^5$ of
\dinstantons. Similarly when both $u$ and $\bu$ are large we should
recover the D-instanton in flat space. This may be checked from the
asymptotic expansions.

In the next section, we will discuss aspects of holography in this
space-time. The boundary will be taken at $u = \infty$. 
The final results should reproduce the
$AdS$ results in the $ka << 1$ limit. It is easy to see that in
the limit $(ka) << 1$ with $y = k/u$ fixed the solutions reduce 
to the solutions for the $AdS_5$ problem. However the holographic
correspondence we are interested in involves taking $u = \infty$
first, before taking the $(ka) << 1$ limit. This latter limit
is nontrivial.

One comment is worth making at this point. We saw in section 2 that the 
D-instanton solution is given by the Greens function for \fourteen. 
The metric which appears in \fourteen\ is the full three brane metric. Thus
\exform\  also provides the solution for the D-instanton in the full D3 
brane geometry
(with no two -form fields turned on). Although we will not do so here,
using \defu\ \exform\ can be expressed in terms of the radial coordinate $r$
in a straightforward manner. 

\newsec{Supersymmetry of the solution}

The type IIB theory has thirty two supersymmetries.
In this section we will consider a solution of the kind \fourt\ \four\ 
and show that it preserves sixteen of the thirty two supersymmetries. 
For definiteness we will consider the case where $\gamma=1$ which is 
the choice made  in this paper. 
The supersymmetry variations of the dilatino and the gravitino
are given by
\eqn\dili{\delta \lambda= {-1 \over 2 \tau_2}({\tau^* -i \over \tau +i})
\Gamma^\mu~\partial_\mu\tau (\eta_1 -i \eta_2)  - {i \over 24} \Gamma^{MNP} 
(\eta_1 + i \eta_2)  G_{MNP}}
\eqn\gravi{\eqalign{&\delta \psi_\mu=\partial_\mu(\eta_1 + i \eta_2)
+{1 \over 4} 
\omega_\mu^{ab}\Gamma^a \Gamma^b (\eta_1+i\eta_2) 
+ {i \over 8 \tau_2 }[({\tau-i \over \tau^*-i})\partial_\mu\tau^*
+ {\rm c.c. }] (\eta_1+i\eta_2) \cr
&~~~ \quad  + {i \over 480} \Gamma^{\mu_1 \cdots \mu_5} \Gamma_\mu 
F_{\mu_1 \cdots \mu_5} (\eta_1+i \eta_2) \cr
&~~~ \quad - i{1 \over 96}(\Gamma_\mu^{NPQ}G_{NPQ} -9 \Gamma^{NP}
G_{MNP}) (\eta_2+i\eta_1)}}
where 
\eqn\defau{\tau= \tau_1 + i \tau_2 = \tau_1 + i e^{-\phi}}
\eqn\defG{G_{MNP}={i \sqrt{\tau_2}}~({|1-i\tau | \over {\tau_2} (1 -i \tau)})~
(H_{RR}-\tau H)_{MNP}. }
$\tau_1$ and $H_{RR}$ are related to the fields used in this paper by 
the relations:
\eqn\relz{\tau_1=-i\chi}
\eqn\defhrr{H_{RR}=-i{\tilde H}.}

Thus, from \fourt\ and ($\gamma=1$) we have
\eqn\rela{\tau_2-\tau_1=c}
and
\eqn\relb{H_{RR}-(\tau_2-\tau_1)H=0.}

\dili\ and \gravi\ are supersymmetry transformations in 
Minkowski space. In going over to Euclidean space we 
follow \ref\gibbons{G. W. Gibbons, M. B. 
Green and M. Perry, Phys. Lett. B370 (1996), 37}
and replace the usual algebra of complex numbers generated 
over real numbers  by $(1,i)$ with "hyperbolic" complex numbers 
generated over real numbers by $(1,e)$ with $e^2=1$. 
Thus $(\tau_1 + i \tau_2) \rightarrow (\tau_1 + e \tau_2)$,
$\eta = \eta_1 + i\eta_2 \rightarrow \eta_1+e \eta_2$
etc.
Then using the fact that $e^2=1$, \rela\ and \relb\ it is easy to 
verify  that the dilatino variation is zero if $\eta_1$ and $\eta_2$
 satisfies the relation 
\eqn\relc{\eta_1 = \eta_2.}

As far as the gravitino variation is concerned we
find from \rela\ \relb\ and \relc\ that the  third term involving 
the derivative of $\tau$ and its complex conjugate vanish. 
Also the last term involving $G_{MNP}$ vanishes. 
Thus the gravitino variation becames:
\eqn\gravittwo{\delta\psi_\mu=\partial_\mu (\eta_1 + e \eta_2) 
+{1 \over 4} 
\omega_\mu^{ab}\Gamma^a \Gamma^b (\eta_1+e \eta_2)+
{e \over 480} \Gamma^{\mu_1 \cdots \mu_5}
\Gamma_\mu F_{\mu_1 \cdots \mu_5} (\eta_1+e \eta_2)}

We argued in section 2 that for the ansatz \fourt\ \four\ the metric and five
form field strengths are independent of the two form fields and 
take the same form that they have in the D3 brane solution without any
two-form expectation values. With this in mind we see that the 
\gravittwo\ is exactly the same as for D3 branes without any 
additional two-form fields turned on. It follows then that the solution
\fourt\ \four\  preserve sixteen supersymmetries. 
It also follows that the spinors $\eta_1,\eta_2$ which satisfy \gravittwo\
and \relc\ are the same as in the case of D3 branes without the 
two-form.

\newsec{Holography}

D- instantons in $AdS_5$ were studied in \dinstantons\ and shown to
correspond to instantons in the Yang Mills theory. One expects the
relationship to be more general and in particular to hold in the
non-commutative Yang Mills theory as well. In this section we will
study the correspondence further by calculating the action and 
correlation functions in the background of the D-instanton solution
determined above and relating them to the corresponding quantities for 
an instanton in the Yang Mills theory. 
Our calculations will be different from the case
of $AdS_5$ space in one important way.  For the geometry under
consideration here, the Einstein metric eq. \eleven\ (to which the
dilaton couples for example) asymptotically becames flat 10
dimensional spacetime.  Thus in relating the D instanton to the Yang
Mills instanton we will neccessarily have to deal with some of the
difficulties of holography in flat space.
 
Below, we  denote the supergravity solution described in Maldacena and Russo 
(with a antiself dual B field turned on ) as the background geometry. 

The D instanton solution in this background  is given by: 
\eqn\inst{e^{\bphi} = C_1 \int [d^4k]~e^{-ik(x-\bx)}
~A^2(k,a)~\psi_k^{(1)}(\nu,\bu)~\psi_k^{(2)}(\nu,u)~~~~~~~u > \bu}
The axion is related to the dilaton  through the condition
\eqn\cond{ e^{-\phi} +i\chi = {1 \over {\hat g}},} 
and the metric and other three-form fields are the same
as in the background solution. 

To begin calculating in the supergravity theory we start with 
the supergravity action which is :
\eqn\pfour{S = {1\over 4 \kappa_{10}^2}\int d^{10}x {\sqrt{g}}[
(\nabla \phi)^2 + e^{2\phi}(\nabla \chi)^2] + S_B + \cdots }
The bulk action is well known. $S_B$ above denotes a boundary term
which is  a little subtle to determine  but important since it plays a crucial
role in holography. For flat space the boundary term was determined 
in \ref\greengutperle{M. Green and M. Gutperle, hep-th/9701093} , by 
working in terms of a eight form gauge potential  dual
to the axion, to be:
\eqn\boundary{S_B=-{1 \over 2 \kappa_{10}^2} \int dS^n e^{2 \phi}
\chi \partial_n \chi.}
 \greengutperle\  showed that \boundary\  gave the correct action
for a D-instanton. One can also show that \boundary\  gives the correct action
for a D-instanton in AdS space. Carrying out a duality transformation to 
the eight form is not straightforward in the present context
since the three form field strenghts etc. are turned on as well. Even so 
we will that the boundary term \boundary\ is unchanged here as well
\foot{Intutively one would expect this boundary term because asymptotically
the three form gauge potentials go to zero and one can at least in that 
region dualise the axion field.}. As we will see below, this choice 
yields sensible answers  at least for small perturbations about the 
background geometry. 

\leftline{ \bf 5.2 The instanton action}
One can now calculate the action for the instanton solution, \exform. 
 From \pfour\  and \cond\ we see that the contribution from the 
bulk action vanishes. As was discussed in \greengutperle\
 since the  D-instanton is a 
singular solution we need to consider two boundaries.
One, as expected, is at large $u$, where the Einstein metric is asymptotically
flat; the other is an infinitesimal  surface  surrounding the location
of the D-instanton. 
Consider first the boundary at infinity. Since the background
geometry is translationally invariant along the branes, only the
 zero momentum piece of the instanton solution \exform\  contributes
to the boundary action. We denote the background as $(\phi_0, \chi_0)$ and 
write the dilaton as $e^\phi=e^{\phi_0}+e^\bphi$ and $\chi=\chi_0+\chi_I$. Then using the 
fact that the full solution and the background satisfy \cond\ the leading
dependence on the D-instanton in the  
boundary action (i.e. after subtracting an infinite contribution from the
background) is : 
\eqn\bca{\Delta S_B= {-i \over 2 \kappa_{10}^2}\int dS^n~( \chi_0 \partial_n
e^{\bphi} + \chi_I \partial_n e^{\phi_0}) }
Using the zero momentum piece from \exform and the condition \cond\
one finds that this vanishes. 
That leaves finally the second boundary: an infinitesimal surface
around the location of the D-instanton. In this case one gets that the relevant
term which contributes is :
\eqn\bccont{\Delta S_B= {-i \over 2 \kappa_{10}^2} \int dS^n \chi 
\partial_n e^\bphi.}
Since $e^\bphi$ satisfies the \fourteen\ it diverges at the location
of the D-instanton. Thus from \cond\ we see that $i\chi \rightarrow \ {1 \over 
{\hat g}}$. Substituting in \bccont\ and using the equation satisfied
by $e^\bphi$ (equation \fourteen), we finally get the action of the 
D-instanton to be 
\eqn\finalac{S= {2\pi \over {\hat g}}}

This result   agrees with the action  of an instanton in the 
noncommutative Yang Mills theory. Note that, in the 
decoupling limit of \seibergwitten\  ${\hat g}$
is the open string coupling. This  is related 
to the Yang Mills coupling by
\eqn\rone{4 \pi {\hat g} = g_{YM}^2}  
(this can be shown for example from the 
Born Infeld action).
Thus \finalac\ yields,  
\eqn\rtwo{S=8 \pi^2/g_{YM}^2.}
This is in fact the 
classical action of an instanton in the non-commutative Yang Mills theory,
as shown e.g. in \ref\aska{A. Astashkevich, N. Nekrasov and A. Schwarz,
hep-th/9810147.}
Further, since
 the instanton is a BPS state we expect its action to not get any quantum 
corrections and is therefore expected to agree with a 
supergravity computation.

It is  worth pausing to emphasise that the calculation above 
in type II  closed string theory yielded an answer in terms of 
the open string coupling constant ${\hat g}$. In fact ${\hat g}$  
is related to the closed string coupling $g_s$ by the relation 
\defhg\  involving the non-commutativity
parameter explicitly. 
 By way of contrast we note that had we not taken 
the scaling limit of the solution \elevena\   an analysis 
along the above lines ( with the same boundary term  )
shows that the action of the D-instanton is  $S=2 \pi /g_s$.
The agreement between the D-instanton action \finalac\  and 
the Yang Mills instanton obtained above provides additional evidence that 
the scaling limit \eleven\  accurately implements the decoupling limit
described in \seibergwitten.  

\leftline{\bf 5.3 One Point Functions}

As mentioned above, we would like to use the D instanton solution to
 calculate expectation values for operators in the instanton
 background of the Yang Mills theory.  Three steps
 are involved determining the expecation values of operators in the
 Yang Mills theory from a supergravity solution.  First, one
 determines the normal modes for small fluctuations about the
 background solution. Next, one computes the change in action to
 linear order in the small fluctuations. Finally, one determines the
 operator in the Yang Mills theory which corresponds to the supergravity
 normal mode and relates the change in the action to the one point
 function of the corresponding operator in the Yang Mills theory.
 
Here we will focus on one normal mode which satisfies the condition
\eqn\nmode{\delta \chi + i e^{-\phi}\delta \phi =0.}
The discussion leading to the instanton solution in section 2
shows that a perturbation  \nmode\ satisfies the 
equations of motion if 
\eqn\nmodeb{\nabla^2(e^{\phi} \delta \phi) =0.}
Solutions to this equation behave asymptotically, for large $u$,  like
\eqn\asform{e^{\phi} \delta \phi \simeq \int { d^4k \over (2 \pi)^4}
C(\vec k)   u^{-5/2} 
[e^{ka^2u} + B(ka) e^{-ka^2u}]} 

In computing the one point function we first compute the change in the 
action for a perturbation about the background to second order. 
For this purpose it is useful to remember that when \nmode\ is met, 
the full solution, consisting of background and perturbation, satisfies
\cond. Thus the bulk contribution vanishes and the change in the action
arises entierly from the boundary term \boundary.  Expanding this to second 
order gives \foot{The one point function of this peturbation vanishes
and this can be shown in exact analogy with the discussion of the instanton
action above.} 
a surface term of the form:
\eqn\secor{S={1 \over 2 \kappa_{10}^2} \int dS^n~ e^{-2 \phi_0}~(
e^\phi \Delta \phi)  ~ \partial_n(
e^\phi \Delta \phi). }
The surface integral is at fixed value of $u=u_0$.
Now we substitute 
\eqn\defD{\Delta \phi = \delta \phi + \phi_I,}
where $\phi_I$ denotes the instanton background,
\eqn\defHH{e^{\phi_0} + e^{\bphi} = e^{\phi_0 + \phi_I}}
 and $\delta \phi$ is given by 
\asform.  
This yields:
\eqn\onep{S = ({5 \pi \over u_0}) e^{-2\phi_0} \int {d^4k \over 
(2\pi)^4} C(\vec{k})A^2(k,a) {1\over \sqrt{ka^2}} \psi_k^{(1)}(\nu,\bar{u})}
and a one-point function 
\eqn\defcone{<O(\vec k)>={\partial S \over \partial C(\vec k)} = 
({5\pi \over u_0}) e^{-2\phi_0}
~ {A^2(k,a) \over (2 \pi)^4} ~{1\over \sqrt{ka^2}}~
\psi_k^{(1)}(\nu,\bar{u})}

In order to relate this result to the gauge theory we need to decide
what is the normalisation of the operator in the Yang Mills theory
which couples to the perturbation.  Unlike $AdS$ space, here there is
considerable ambiguity in this matter.  This issue was explored in
\maldarusso\ for a massless scalar field where it was found that a
momentum dependent wave function renormalisation needed to be done to
extract sensible answers. A similar renormalisation is required here
as well \foot{This is clear if we consider a two point function of the
perturbation above the background geometry.}.  We have not been able to
implement a satisfactory renormalisation scheme at the moment. The one
which is most natural in the present context corresponds to a momentum
dependent wave function renormalisation and therefore does not
correpond to a choice of local counter terms.  Below we will
choose a renormalised
operator whose one point function agrees at low
momenta, and for a large instanton, with expectations from the
ordinary gauge theory.  The perturbation \nmode\ should couple to the
operator $TrF^2 + TrF{\tilde F}$ at low momenta. Now the function
$A^2(k,a)$ behaves for low momentum as:
\eqn\lowa{A^2(k,a) = c_1 (ka)^4 + \cdots}

Then consider the renormalised  operator given by:
\eqn\defro{O_R(\vec k) = {u_0 \hg \over 5 c_1} {e^{2\phi_0} \over (ka^2)^{3/2}}
 O(\vec k)}
It's expectation value from \defcone\ 
is
\eqn\emom{<O_R(\vec k)>={\pi \over \hg}~k^2 \psi^{(1)}_k(\nu,\bu)}
When $a\bu << 1$ the function $\psi^{(1)}_k(\nu,\bu) 
\rightarrow {1\over \bu^2} K_\nu (k/\bu)$, and for $(ka) <<1$ we
can replace $\nu \sim 2$. Then \emom\ reduces to
exactly the answer for the one point function in the presence of
a D-instanton in $AdS_5 \times S^5$ \foot{This may be verified by writing
the D-instanton solution as a momentum space integral using the techniques
of this paper and repeating the above calculation in that case.}. This is  
the Fourier transformation of 
{\eqn\insex{{1 \over 4 g_{YM}^2} <Tr F^2(x) + TrF{\tilde F(x)}> = 
{48 \over g_{YM}^2}
 {({1 \over {\bar u}})^4
\over [({1 \over {\bar u}})^2 + |\vec x - \vec x^{'}|]^4},}
which is the profile of an instanton in ordinary Yang-Mills theory.

Thus when the D-instanton is located deep in the bulk and we look at
the long wavelength modes of the solution, the parameter $1/{\bar u}$
in \emom\ plays the role of the instanton scale size as expected.
It would be interesting to examine the departures from this
low momentum behavior, as well as the predictions for the noncommutative
Yang-Mills instantons of small size.
 
\leftline{\bf 5.4 Operators in the Gauge Theory}

In the usual $AdS-CFT$ correspondence, the dilaton $\phi$ couples to the
operator as 
\eqn\qone{\int e^{-\phi} {\rm Tr}(F_{ij}F^{ij})}
while the axion couples as
\eqn\qtwo{\int \chi {\rm Tr}(F_{ij}{\tilde F}^{ij})}
where ${\tilde F}$ stands for the dual in the four dimensional sense.
One might think that to get the operators for noncommutative
Yang-Mills theory one simply needs to replace ordinary gauge fields
$F$ by nonocommutative gauge fields ${\hat F}$ and odrinary products
by star products in the above relations. 
However this cannot be correct. Unlike in ordinary Yang -Mills theory,
objects like $Tr{\hat F} * {\hat F}$ are not gauge invariant though their
integrals are \foot{S.T. acknowledges a  discussion with A. Hashimoto 
on this issue.}.   As a result, while the operator
${\rm Tr} {\hat F_{ij}}*{\hat F_{kl}}$ can be coupled to 
supergravity fields which are constant along the brane (as for example 
in the form of the action for noncommutative
Yang-Mills), one cannot couple them to nonconstant supergravity
fields. However nontrivial effects of noncommutativity appear when
there is momentum along the brane directions.

It is possible to write down gauge invariant operators order by order
in the noncommutavity parameter and couple supergravity fields, but
one suspects that there must be a more natural and simpler principle
here. In a sense noncommutative gauge theories have a ``nonabelian''
nature when fields are regarded to have their momenta as group indices.
Thus the natural invariant objects are in fact integrals over the
brane rather than local quantities.

In any case, we believe that our supergravity results should act as a
guide to the correct coupling of noncommutative gauge field with
closed string fields.

\newsec{Acknowledgements}

We would like to thank Sunil Mukhi for discussions and collaboration
at the early stages of this work and A. Hashimoto, M. Krogh,
S. Mahapatra, P. Ramadevi and D. Tong for
discussions. S.R.D would like to thank the participants of the
Workshop on String Theory at MRI, Allahabad for discussions and
questions which clarified many points.
S.P.T. would like to acknowledge the hospitality of the ITP,
Santa Barbara, where some of this work was completed as a participant
in the workshop on  Supersymmetric  Gauge Theories, 
supported by the NSF grants, Phy 94-07194 and Phy 97-22022. 

\newsec{Appendix I : Supergravity equations.}

The equations of motion for the bosonic fields in IIB supergravity
in ten dimensions are, in terms of the einstein metric
\eqn\onea{R_{\mu\nu} - {1\over 2}g_{\mu\nu}R = T_{\mu\nu}^\phi
+ e^{-\phi} T_{\mu\nu}^B + e^\phi T_{\mu\nu}^\tB +e^{2\phi}
T_{\mu\nu}^\chi + T_{\mu\nu}^F}
\eqn\oneb{{1\over \sqg}\partial_\mu(\sqg g^{\mu\nu}\partial_\nu \phi)
+{1\over 12}[e^{-\phi}H^2 - e^\phi(\tH - \chi H)^2]
-e^{2\phi}g^{\mu\nu}\partial_\mu\chi\partial_\nu\chi = 0}
\eqn\onec{{1\over \sqg}\partial_\mu
(\sqg g^{\mu\nu}e^{2\phi}\partial_\nu \chi)
+ {1\over 6}(\tH_{\mu\nu\lambda} - \chi H_{\mu\nu\lambda})
H^{\mu\nu\lambda} = 0}
\eqn\oned{{1\over \sqg}\partial_\mu(\sqg e^{-\phi} H^{\mu\nu\lambda})
-{1\over \sqg}\partial_\mu(\sqg e^{\phi}\chi
(\tH^{\mu\nu\lambda} - \chi H^{\mu\nu\lambda}))
+ {2\over 3}F^{\nu\lambda\alpha\beta\gamma}\tH_{\alpha\beta\gamma} = 0}
\eqn\onee{{1\over \sqg}\partial_\mu(\sqg e^{-\phi} (\tH^{\mu\nu\lambda}
- \chi H^{\mu\nu\lambda})) - 
{2\over 3}F^{\nu\lambda\alpha\beta\gamma}H_{\alpha\beta\gamma} = 0}
\eqn\onef{{1\over \sqg}\partial_\mu(\sqg 
F^{\mu\lambda\alpha\beta\gamma}) =0}
The fields are as following. $\phi, \chi$ denote the dilaton
and the axion. $B, \tB$ denote the NS and RR 2-form fields and
$H, \tH$ are the corresponding field strengths which are defined as
\eqn\two{H_{\mu\nu\alpha} = \partial_\mu B_{\nu\alpha}
+ \partial_\nu B_{\alpha\mu} + \partial_\alpha B_{\mu\nu}}
and similarly for $\tH$ in terms of $\tB$. 
$F^{\nu\lambda\alpha\beta\gamma}$ is the self dual 5-form field 
strength. The energy momentum tensors appearing in \onea\ are given
by 
\eqn\three{\eqalign{&T_{\mu\nu}^\phi 
= {1\over 2}[ \partial_\mu\phi\partial_\nu\phi - {1\over 2}g_{\mu\nu}
(\partial \phi)^2] \cr
&T_{\mu\nu}^\chi 
= {1\over 2}[\partial_\mu\chi\partial_\nu\chi - {1\over 2}g_{\mu\nu}
(\partial \chi)^2] \cr
&T_{\mu\nu}^B = {1\over 12}[3 H_{\mu\alpha\beta}H_\nu^{\alpha\beta}
- {1\over 2}g_{\mu\nu} H^2]\cr
&T_{\mu\nu}^\tB = {1\over 12}[3 (\tH - \chi H)_{\mu\alpha\beta}
(\tH - \chi H)_\nu^{\alpha\beta}
- {1\over 2}g_{\mu\nu} (\tH - \chi H)^2]\cr
&T_{\mu\nu}^F = {1\over 480}[5 F_{\mu\alpha_1\alpha_2\alpha_3\alpha_4}
F_\nu^{\alpha_1\alpha_2\alpha_3\alpha_4} - {1\over 2}g_{\mu\nu}F^2]}}

\newsec{Appendix II: Asymptotic forms and the Wronskian}

In this appendix we give the necessary asymptotic forms of the solutions
and various quantities related to these solutions and compute the
Wronskian necessary for normalization of the Green's function.

\subsec{Asymptotic forms}

The asymptotics of $\psi_k^{(i)}$ may be figured out by using the
relation \gubserhashimoto,
\eqn\twoseven{H^{(1)}(\nu,w) = {1\over 2i \sin \pi\nu}
[(\zeta -{1\over \zeta})H^{(1)}(\nu, -w)+
(\zeta -{e^{-2\pi i \nu}\over \zeta})H^{(1)}(\nu, -w)]}
where we have defined
\eqn\twosevena{\zeta = {\phi (-\nu/2)\over \phi(\nu/2)}}
We will also define
\eqn\threeone{\eta = e^{i\pi\nu}}
$\phi(\nu)$ is the function of $\nu$ which appears in the
Floquet form of the solution to Mathieu equation.
The relation \twoseven\ yields the following asymptotics for $w
\rightarrow -\infty (u \rightarrow \infty)$
\eqn\twoeight{\eqalign{& \psi_k^{(1)}(\nu,u)
\rightarrow {1\over u^2}[{i e^{-i\pi\nu}\over \zeta}K_\nu (ka^2u)
+ {\pi \over 2\pi \sin \pi\nu}(\zeta e^{i\pi\nu} -
{e^{-i\pi\nu}\over \zeta})I_\nu(ka^2u)]\cr
&\psi_k^{(2)}(\nu,u) \rightarrow {1\over u^2} K_\nu(ka^2 u)}}
and similarly for $w \rightarrow
\infty (u \rightarrow 0)$
\eqn\twonine{\eqalign{&\psi_k^{(1)}(\nu,u) \rightarrow
{1\over u^2} K_\nu({k\over u})\cr
& \psi^{(2)}(\nu,u)
\rightarrow {1\over u^2}[{i e^{-i\pi\nu}\over \zeta}
K_\nu ({k\over u})
+ {\pi \over 2\pi \sin \pi\nu}(\zeta e^{i\pi\nu} -
{e^{-i\pi\nu}\over \zeta})I_\nu({k\over u})]}}

\subsec{The Wronskian}

It follows from standard Strum-Liouville theory 
that  $W(\psi_k^{(1)}, \psi_k^{(2)}) u^5$
is a constant independent of $u$, 
so we can evaluate it for any $u$. In particular
we may evaluate this for $ w \rightarrow -\infty$. 

The Wronskian can be now easily evaluated, giving
\eqn\thirty{W(\psi_k^{(1)},\psi_k^{(2)}) = {1\over u^5}
[{\pi \over 2\sin \pi\nu}(\eta \zeta -{1\over \eta\zeta})]}
This fixes the function $A^2(ka)$ defined in the text to be
\eqn\threetwoa{A^2(ka) = {2\sin \pi\nu \over \pi}
(\eta \zeta -{1\over \eta\zeta})^{-1}}

\subsec{Power series expansions of various quantities}

The results of \gubserhashimoto\
may be used to obtain expressions for
the various quantities defined above. First note that in our case,
the expansion for $\nu$ ensures that $\eta$ is real. 
Then, as shown in \gubserhashimoto,
the quantity $\zeta$ has to have unit modulus. Let us denote
\eqn\zone{\eta = e^\beta~~~~~~~~\zeta = e^{i\alpha}~~~~~~(\alpha,\beta~~
{\rm real})}
The quantity $\beta$ can be read off from the definition of $\nu$.

Define the quantities
\eqn\ztwo{\eqalign{&E=\zeta - {1\over \zeta}= 2\sin \alpha\cr
&F=\zeta\eta - {1\over \zeta\eta}= 2(\cos\alpha \sinh \beta + 
i \sin\alpha \cosh \beta)\cr
&G=\eta - {1\over \eta} = 2\sinh \beta}}
The unitarity relation of the quantum mechanical problem corresponding to
our basic differential equation
\eqn\zthree{|F|^2=|E|^2+|G|^2}
is now automatically satisfied.

In terms of the quantity
\eqn\zfour{ P = |{G \over F}|^2}
it is straightforward to show that
\eqn\zfive{\sin^2\alpha = \sinh^2\beta({1\over P}-1)}

We can now obtain expressions for $E,F,G$ in terms of $\beta$ and
$P$. $\beta$ has been already given in the expansion for $\nu$ above, 
\eqn\zfoura{\beta = {{\sqrt {5}}\over 3}({ka\over 2})^4
- {7 \over 108 {\sqrt{5}}}({ka\over 2})^8 + \cdots}
while $P$
has been expressed as a power series in \gubserhashimoto. In our
notation this is
\eqn\zfive{P = \pi^2 ({ka\over 2})^8 \sum_{n=0}^\infty b_{n,k}
(ka)^{4n}(\log (ka))^k}
where $b_{n,k}$ are some real coefficients tabulated in \gubserhashimoto.

Of particular interest is the small $ka$ behavior of these quantities.
These are
\eqn\zsix{\eqalign{& E = e_1 + O((ka)^4)\cr
&F= i f_1 + O((ka)^4)\cr
&G = g_a (ka)^4 + O((ka)^8)}}
where $e_1,f_1,g_1$ are real numerical coefficients of order one.
\eqn\zsix{A^2(ka) = {1\over i\pi}{G\over F} \sim (ka)^4}
We will use these results to examine the low momentum components of
the instanton solution.

\subsec{Low energy limit of solutions}

In the limit $(ka) << 1$ at fixed $(au)$ the solutions $\psi_k$ go
over to Bessel functions as well. This may be seen from the relation,
valid for any $w$
\eqn\xone{H^{(1)}(\nu,w) = \sum_{n=-\infty}^\infty {\phi(n + {nu\over 2})
\over \phi({\nu\over 2})} J_n (ika~e^{-w})~H^{(1)}_{n+\nu}(ika~e^w)}
For $(ka) <<1$ with fixed $w$ only the $n=0$ term in the sum
contributes, and we can also set $\nu =2$ in the leading order. 
We finally have, in this limit
\eqn\xtwo{\eqalign{&\psi_k^{(1)}(\nu,u) \rightarrow {1\over u^2}K_2(k/u)\cr
&\psi_k^{(2)}(\nu,u) \rightarrow {1\over u^2}K_2(ka^2u)}}

\listrefs
\end